# A first-principles-based study of the thermodynamics of competing low-energy states in correlated materials: Example of cuprates

Robert S. Markiewicz,[1,*,†] Yubo Zhang,[2,*] Christopher Lane,[1] Bernardo Barbiellini,[3,1] Jianwei Sun[2,†] and Arun Bansil,[1,†]

[1]*Department of Physics, Northeastern University, Boston, MA 02115, USA*
[2]*Department of Physics and Engineering Physics, Tulane University, New Orleans, LA 70118, USA*
[3]*Department of Physics, School of Engineering Science, Lappeenranta University of Technology, FI-53851 Lappeenranta, Finland*

[†]To whom correspondence should be addressed; E-mails r.markiewicz@neu.edu, ar.bansil@neu.edu, jsun@tulane.edu

*: These authors contributed equally

**ABSTRACT**

We demonstrate how first-principles calculations of many competing low-energy states of a correlated material, here a cuprate, can be used to develop a thermodynamic model of Mott and pseudogap transitions in terms of magnetic short-range order. Mott physics is found in this picture to be driven by an unbinding of the antiphase domain walls, while the pseudogap phenomenon represents local moment formation. We provide explanations for nematicity and Fermi arc formation, and find a striking correspondence with many-body perturbation theory predictions.

**Popular Summary**

By analyzing the competition between many first-principles-derived low-energy phases in $YBCO_6$ and $YBCO_7$ as exemplar complex correlated materials, we show how the resulting thermodynamic picture can capture successfully many salient features of the physics of the cuprates, including Mott and pseudogap transitions and nematicity..

**I. INTRODUCTION**

It is becoming increasingly clear that the physics of many strongly-correlated materials is controlled by strong competition among multiple phases. For instance, the pseudogap phase of cuprates hosts a bewildering variety of competing phases, including the antiferromagnetic (AFM), charge-density wave (CDW), spin-density wave (SDW), and the stripe or nematic phase. This behavior has been labelled as 'intertwined orders,' see Ref. [1] for a critical review of the related theories. Here we show how a thermodynamic analysis can be used to gain insight into first-principles results on competing phases in complex materials to build a new model, which can capture many anomalous features of the physics of the curpates as an exemplar complex material.

Our analysis is based on the large number of competing phases that can be obtained through first-principles computations on undoped $YBa_2Cu_3O_6$ ($YBCO_6$) and near-optimally-doped $YBa_2Cu_3O_7$ ($YBCO_7$) using advanced exchange-correlation functionals.[2] The set of phases used here involves magnetic and stripe states, which are nearly degenerate with the AFM state in $YBCO_7$, suggestive of a pseudogap phase dominated by fluctuations, along with similar stripe phases in $YBCO_6$. Our thermodynamic model suggests that down to temperatures close to the pseudogap and Mott transitions, the electrons form a strongly-fluctuating paramagnetic state (spin liquid) with only short-range magnetic order.

It has recently been noted that the SCAN exchange-correlation functional tends to find a 'too large' magnetic moment in the ground state of many materials[3,4]. A similar criticism could be leveled against Ref. [2], which finds a stripe phase ground state in YBCO7, except that recent Hubbard model calculations find a similar ground state for doped cuprates[5], and there is strong experimental evidence for fluctuating magnetic moments in heavily doped cuprates[6,7]. The present paper will demonstrate that large moments are an asset and not a liability in trying to understand pseudogap physics in cuprates.

## II. PATCH MAPS

The competing fluctuations can be organized into a temperature- and doping-dependent patch map where the different patches correspond to the various competing phases with a probability that is controlled by an underlying partition function. We will refer to this model of fluctuations as the Density-Functional Theory-based Multiphase Approximation (DFT-MPA). The high electronic mobility and weak coupling to the lattice suggest that these phases can peacefully coexist on a sufficiently nanoscopic scale. In what follows, we will demonstrate that the DFT-MPA provides key insights into a number of features of the pseudogap, including nematicity, CDW fluctuations, and Fermi arcs.

We will define a patch as a minimal fluctuation of a given phase, where the minimum size of a fluctuation is assumed to be one unit cell. These patches are assumed to have a canonical distribution. Specifically, we assume that (1) the probability that a patch of state $i$ would appear at temperature T is proportional to a Boltzmann factor $P_i = exp(-n_i E_{0i}/k_B T)$, where $n_i$ is the number of copper atoms per unit cell, and $E_{0i}$ is the energy per planar copper of the given phase. The choice of a single unit cell as minimum fluctuation size is consistent with the result of a recent machine learning study of scanning tunneling microscopy (STM) maps[8]. (2) By shifting the boundaries of the unit cell, $n_i$ independent patterns of atoms can be formed, and these should be treated as independent fluctuations. Thus, to form the partition function $Z$, $P_i$ should be weighted by the state degeneracy $n_i$, so $Z = \sum_i Z_i$, where $Z_i = n_i P_i$. (3) To form a patch map that can be compared to STM images, we need to find the fraction of the total area occupied by phase $i$, which is proportional to $Z_i$ multiplied by the unit cell size, $A_i = n_i Z_i$, normalized by $A_Z = \sum_i A_i$. (4) Finally, while in the orthorhombic $YBCO_7$ structure, stripes along $a$ or $b$ axes have different energies, they are degenerate for tetragonal $YBCO_6$. Hence, in the latter case, we double $A_i$ for each stripe to account for the two axes. Further details of the model are discussed in Appendix I, including a discussion of the free energy and mixing entropy.

The patch method works because each phase is characterized by its total energy, which DFT can find exactly. For $YBCO_7$, there are 21 distinct phases found on a single $CuO_2$ plane in Ref. [2]. In addition to the G-AFM (or $(\pi,\pi,\pi)$) order, there are 20 stripe phases, which are AFM phases with periodic arrays of charged antiphase boundaries (domain walls), including 10 site-centered stripe phases, where the charged stripe is centered on a single row of Cu, and 10 bond-centered stripe phases, in which the charge stripe consists of two rows of Cu, with ferromagnetic coupling between them. There are one dimensional (1D) stripes running perpendicular to either *a*- or *b*- axes, as well as 2D stripes. [We omit three other phases that lie at too high an energy to be relevant in low-energy physics, including two in-plane ferromagnetic phases and the nonmagnetic Fermi liquid phase found in DFT calculations.]

Figure 1 is a schematic rendering of a single-layer model of the YBCO$_7$ patch map at three different temperatures. As temperature T→0, all patches represent the ground state, vertical stripes with charge periodicity $P_c$=4 (in units of the lattice constant a), but already at T=50K several patches represent other phases. [Note that the color bar identifies the different phases.] In the bottom row only these ground-state patches are shown, in black, and it is clear that the ground state only percolates across the field of view up to some temperature below 200K. In Section III we discuss this percolation transition and illustrate several other applications of patch maps to experiment.

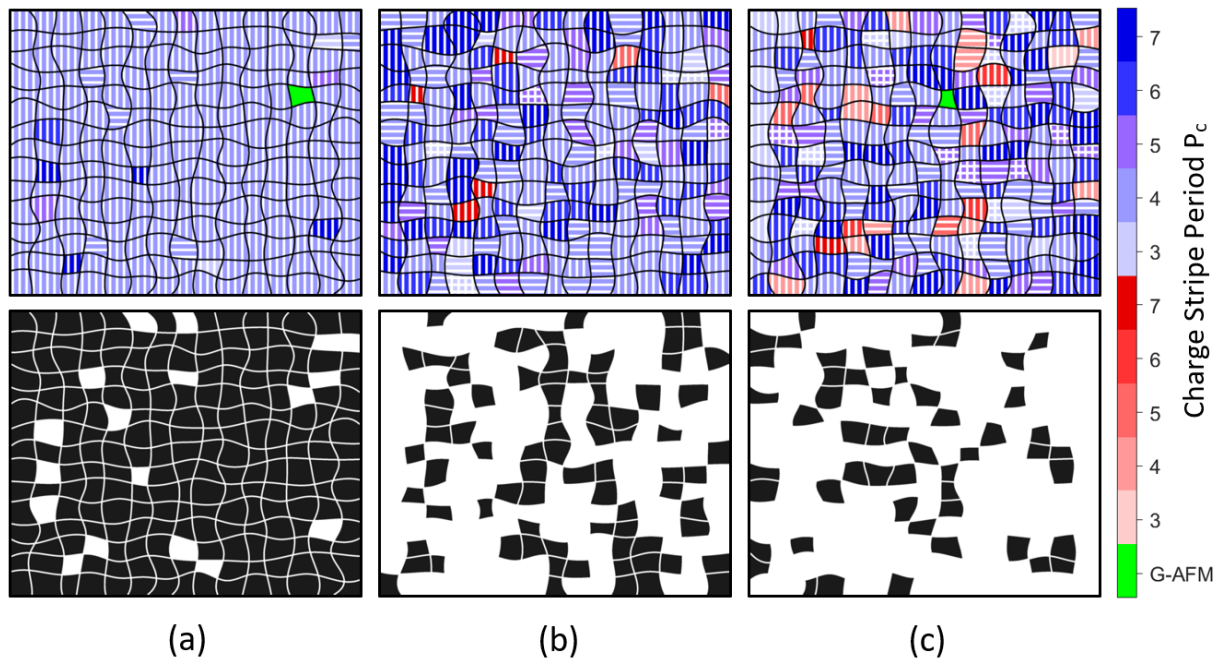

(a)  (b)  (c)

Figure 1. Schematic rendering of YBCO$_7$ patch maps at T = 50 (a), 200 (b), and 400K (c). The 21 phases include the ground-state-AFM (G-AFM) phase (green), 10 bond-centered stripe phases (blue), and 10 site-centered stripe phases (red). Horizontal, vertical, or two-dimensional (2D) stripes are indicated by hatching in the patches, while the shades of red or blue indicate charge-stripe periodicity $P_C$, as indicated in the color bars. Bottom row shows only the ground state patches, illustrating the percolation crossover between a and b.

The construction of the fluctuation partition function is completely generic, relying only on DFT data, and can be applied to any correlated system. Going beyond these results is material specific, and requires determining the organizing principle of the competing phases. In Section IV it is shown that in YBCO the phases are dominated by topological defects -- antiphase boundaries or stripes -- and the Mott and pseudogap transitions constitute a form of defect unbinding, with clear analogies to a number of other physical problems, including vortices in liquid He[9], dislocations in lattices[10] and

Berezhinskii-Kosterlitz-Thouless (BKT) physics of the unbinding of zero-dimensional topological defects in 2D superconductors and x-y magnets[11].

## III. COMPARISON WITH EXPERIMENTS

### A. Patch Maps and STM Images

Since YBCO$_6$ and YBCO$_7$ lack intrinsic disorder, the patch fluctuations should be dynamic, and this is consistent with the observation of local moments by resonant inelastic x-ray scattering (RIXS), but not by slower local probes[6,7,12]. However, since the patches are charged, pinning by impurity disorder in doped materials is also possible. Here we explore the possibility that patch maps pinned near surfaces are observable in STM studies.

Scanning tunneling microscopy (STM) studies of incommensurate CDWs in cuprates find commensurate patches separated by dislocations [13], which bear a striking resemblance to our patch maps in Fig. 1. Indeed, in a recent AI-based study to decode STM images in cuprates[8], only a single pattern could be consistently resolved – a $P_c=4$ charge stripe, consistent with the dominant mode in Fig. 1(a). Indeed, the presence of stripes with many different periodicities at a single doping[2,5] was unexpected, and we suggest that they may be controlled by an incommensurate CDW. This idea is further discussed in Appendix II.

The patch maps of Figure 1 can also be compared with superconducting gap maps found in tunneling studies of cuprates. In the latter, the domain size is set by the superconducting coherence length, ~3nm, which is larger than most of the magnetic and stripe unit cells. It was found that different superconducting gap values correspond to different Fermi surface sizes – i.e., different dopings[14,15]. This is consistent with the finding[2] that different patch phases have different effective dopings. Indeed, it is likely that superconductivity [and other exotic orders in various correlated materials] can leverage the AFM-CDW competition, by tipping the balance between the various patches. One note of caution: the spread in doping in experimental gap maps is considerably greater[16] than found in Ref. [2]. This is likely due to nanoscale phase separation or impurity effects, which would be important in cuprates but are beyond the scope of this study.

Our results for YBCO$_7$ suggest that the stripes are predominantly one-dimensional, consistent with the analysis of Ref. [1]. However, STM studies at high magnetic fields find features[17] that resemble the charge distributions of our 2D stripes[2]. Experimentally, these stripes are pinned to magnetic vortices, which may tend to favor a 2D pattern. This is also consistent with the usual interpretation of the Fermi surface responsible for high-field quantum oscillations.[18]

### B. Patch Percolation in YBCO$_6$ as Mott Physics

It has been claimed that it is impossible to describe a Mott insulator based on DFT, since in DFT the magnetic gap must be accompanied by a long-range superlattice. However, the patch model requires only short-range order, and we find that entropic effects drive a percolation crossover in undoped YBCO$_6$, from a predominantly G-AFM order (green line in Fig. 2(a)) at low T to a disordered paramagnetic liquid – a Mott-like phase – at high-T, Fig. 2(a). In this paper we carefully distinguish a *paramagnetic* liquid, where the individual patches have local moments on Cu, from the *nonmagnetic* phase found in earlier DFT calculations, where local moments are absent. Notably, with no free parameters we find that the point where the AFM probability $a_1$ is ~0.5 -- the 2D percolation crossover -- falls close to the Neel temperature for YBCO$_6$.

In addition to the AFM order, there is a series of stripe phases[2] (blue lines), with different charge periodicity $P_c$. In Fig. 2(a), the curves represent the individual $a_i = A_i/A_z$ as a function of T, with the numbers representing $P_c$ of the individual stripes. It is the gradual growth of these stripe fluctuations at higher T that destroys the long-range Neel order.

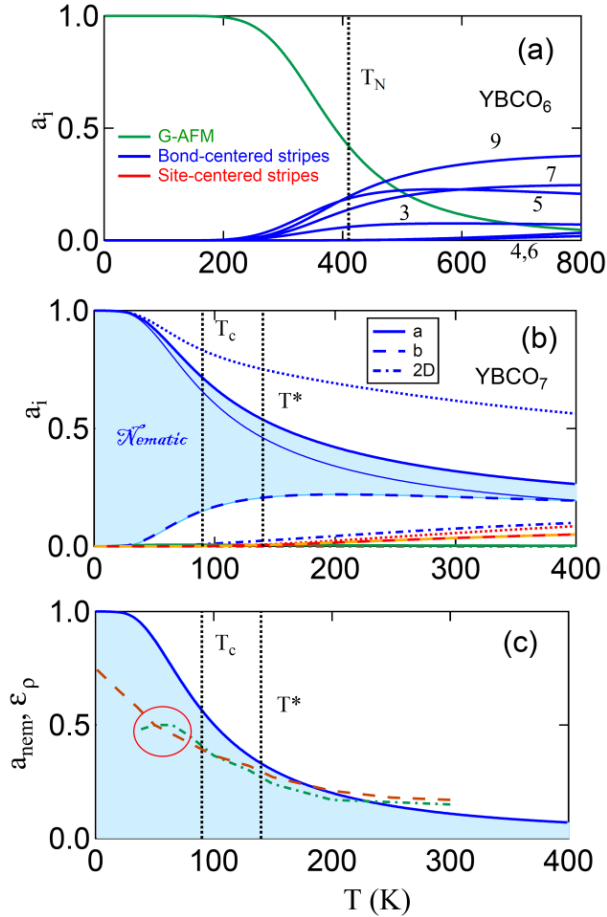

Figure 2. (a) Individual patch weights in YBCO$_6$ vs T. (b) Sums of patch weights for different classes of phases in YBCO$_7$. Blue curves represent bond-centered stripes, red curves site-centered stripes, and the green curve is G-AFM. Solid and dotted lines are stripes modulated along the *a*-axis, dashed lines are modulated along the *b*-axis, and dot-dashed lines are 2D stripes. Since more stripes were found for *a*-axis modulation, the dotted line represents the sum of all 5 phases, whereas the solid line represents the 3 phases also found along *b*, to give a better estimate of nematicity effects. Vertical dotted lines represent the experimental pseudogap temperature T* and superconducting temperature $T_c$. (c) Comparison of calculated nematicity to observed conductivity anisotropy at $\delta$ =0.35 (red dashed curve) and 0.45 (green dot-dashed curve) [19]. The dip at lowest T (circled region) is due to competition with superconductivity.

### C. Patch Percolation in YBCO$_7$ as Pseudogap Transition

The phase diagram of YBCO$_7$, Figure 2(b), is richer than that of YBCO$_6$. The competing in-plane orders include the G-AFM order (green curve), 10 site-centered stripe phases (red curves), and 10 bond-centered stripe phases (blue curves). Rather than plot them individually, as in Fig. 1, we group them as all the stripes modulated along the *a*-axis (dotted lines), or *b*-axis (dashed lines), or along both axes (dot-dashed lines). The patch maps of Fig. 1 illustrate the individual phases at T = 50, 200, or 400K.

Figure 2(b) shows that long-range stripe order with $P_c$=4 ($a_1 \rightarrow 1$) (thin blue solid line) only sets in at $T \rightarrow 0$, while $a_1 \sim 0.5$ close to the pseudogap temperature $T^*$=140K. Thus, the present model can describe both the Mott crossover in YBCO$_6$ and the pseudogap crossover in YBCO$_7$, with no free parameters, as both due to the onset of domain wall fluctuations with increasing T (similar to the BKT transition[11]). In both cases, the high-T phase is a paramagnetic metal that is not related to the nonmagnetic phase found in previous DFT calculations. In Section IV we discuss why the domain wall energy decreases with doping.

### D. Origins of Nematicity

We note from Fig. 2(b) that there are considerably more stripe fluctuations modulated along the *a*-axis than along the *b*-axis, leading to strong nematicity. To provide a quantitative measure of nematicity, we must correct for the fact that 5 *a*-stripes were studied, but only 3 *b*-stripes – the thick blue line represents the corresponding three, and the blue shaded region is the resulting stripe excess, or nematicity $Z_{nem}$. This is plotted in Figure 2(c) and compared to an experimental measure of nematicity, $\varepsilon_\rho = (\rho_a - \rho_b)/2\rho_b$, where $\rho_a$ and $\rho_b$ are the measured anisotropic resistivities in YBCO$_{6+\delta}$.[19]. This comparison can only be approximate, for two reasons. First, for larger oxygen $\delta$ the stripe resistivity is hidden by superconductivity or the chain-layer conductivity, so we can only compare YBCO$_7$ to the stripe-dominated results at $\delta$=0.35 and 0.45[19]. Second, the comparison assumes that the anisotropy of a given stripe is T-independent, and the observed anisotropy arises solely from the reorientation of stripes with increasing T. Despite these limitations, agreement with theory is quite reasonable.

Returning to STM experiments, it was found that both nematic and CDW intensity are maximal close to the pseudogap transition temperature[8]. Our results demonstrate that the percolation transitions in both YBCO$_6$ and YBCO$_7$ correspond to the domain wall unbinding transition, thereby explaining the STM results.

### E. Patch Effects in ARPES

There are a number of anomalous effects, which should be strongly impacted by fluctuating short-range order. Here we briefly discuss three.

#### (a) Nonmagnetic dispersion

A remarkable finding in ARPES studies is that, whereas stripe phases are not clearly seen, traces of the nonmagnetic dispersion and Fermi surface can clearly be seen, even down to very low doping[20]. This can be qualitatively understood in a patch model. When any of the ordered phase dispersions is calculated, unfolding it introduces a coherence factor that modulates the intensity, given by the overlap of the ordered wave function and the nonmagnetic wave function[21]. When the dispersions of many phases are averaged over, the resultant weight will be strongest along the bare dispersion (where the overlap is ~ 1), with other features leading only to a featureless broadening. However, since gaps tend to be localized near the Fermi level, the intensity of the ARPES spectral weight should systematically decrease as the Fermi level is approached, reflecting the gap distribution. This spectral weight loss is in good agreement with ARPES experiments.[20]

### (b) Final-state broadening

A second ARPES anomaly is that the effective spectral broadening depends strongly on the photon energy $\omega$ – that is, on the ARPES final state[22]. This is understandable in a patch model, since the spectral broadening depends equally on T and $\omega$, typically in the combination $[(k_BT)^2+(\hbar\omega)^2]$. Extending this reasoning to optical spectra suggests an analogy to the optical spectra of dirty superconductors.

### (c) Fluctuation effects on interlayer hopping and Fermi arcs

Finally, we briefly explore some effects of patch flucuations on the coherence of interlayer hopping. We analyze only a simple model, neglecting stripes and comparing three different interlayer stackings of the $(\pi,\pi)$ AFM phase, comparing parameter values for YBCO$_6$ and YBCO$_7$[2]. In Fig. 3(a) we compare the G-, C-, and C'-AFM orders. The first two correspond to ordering vectors $(\pi,\pi,\pi)$ and $(\pi,\pi,0)$ respectively, while the last has mixed stacking, AFM across the Y layer and ferromagnetic (FM) across the CuO chain layer.

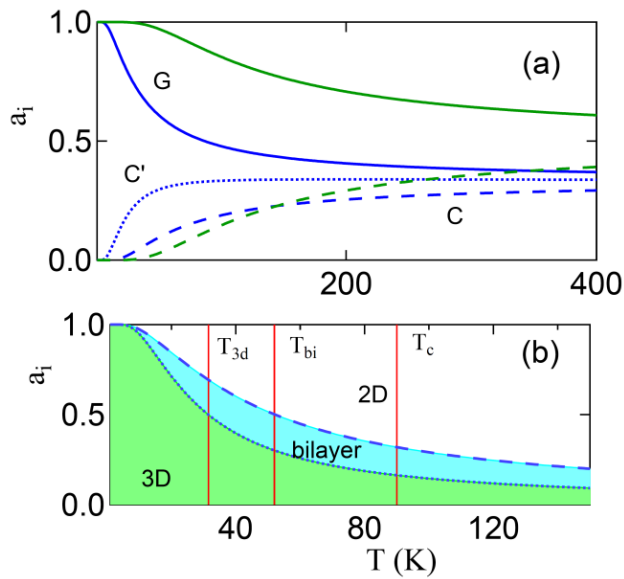

Figure 3. Evolution of interlayer AFM coupling. (a) Interlayer patch weights $a_i$ in YBCO$_6$ (blue lines) and YBCO$_7$ (green lines), comparing three intralayer AEM phases, G-AFM (solid lines), C-AFM (dashed lines), and C'-AFM (dotted line – only found for YBCO$_7$[2]). (b) Interlayer phase diagram, where **3D** (green region) implies interlayer hopping is predominantly within G-AFM phase, and **bilayer** {light-blue region) implies the interlayer coupling across the CuO chain layer has become incoherent. In the white region, all interlayer hopping is incoherent.

Thus, in YBCO$_7$, (blue lines), it can be seen that the weights of the C'-AFM (dotted line) and C-AFM phases (dashed line) grow with increasing *T*. When the weight of C'-AFM order becomes comparable to that of G-AFM order, coherent hopping across the chains is absent, leaving only a bilayer splitting. We plot the difference in weights between the G- and C'-phases as the dotted blue line in Fig. 3(b), defining the point where this line falls to a value 0.5 as $T_{3d}$. We define $T_{bi}$ similarly, from the difference between the C-AFM and G-AFM weights (blue dashed line). Above $T_{bi}$ all coherent interlayer hopping is lost. Similar effects arise in YBCO$_6$, (green lines). For YBCO$_7$, we expect only incoherent interlayer hopping above $T_c$.

Although other explanations are possible[22], the present result suggests a simple explanation for the Fermi arcs found in the pseudogap phase. Figure 4(a) shows that the Fermi surfaces (FSs) expected for the ground state G-AFM order (red curves) have a striking asymmetry, with resulting equal areas. To our knowledge, such FSs have never been seen experimentally in bilayer cuprates. In contrast, the FSs for C-AFM order (blue curves) are symmetric and nested. Figure 4(b) replots the Fermi surfaces with the appropriate AFM coherence factor weights [see Appendix III]. This already provides a partial explanation for the appearance of arcs for either the G- or C—AFM phases. However, fluctuations further enhance the effect, Figure 4(c). In the temperature range above $T_{bi}$, electrons hopping from the G-AFM on one layer will randomly encounter either G- or C- stacking on the next layer, blurring the FSs between the red and blue limits. This will leave the least broadening where the FSs nearly touch, leaving a pattern resembling the Fermi arcs. Cuprate Fermi surfaces are further discussed in Appendix III.

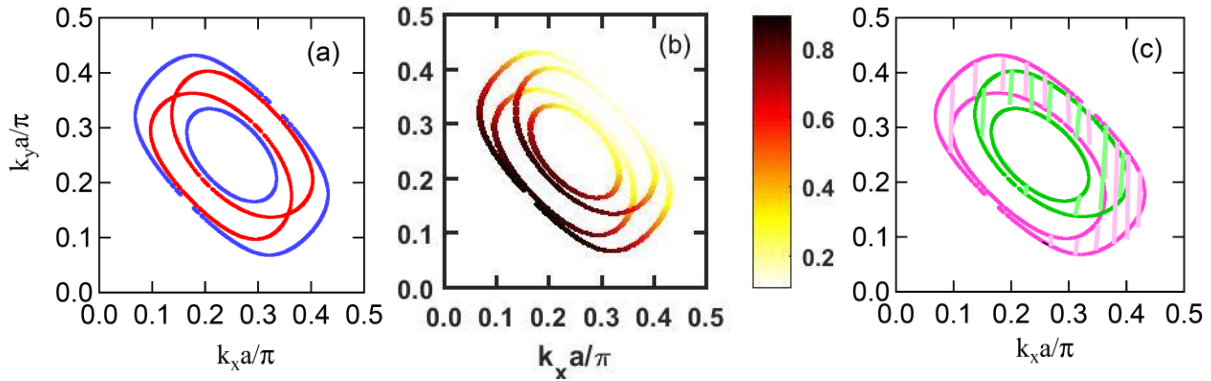

Figure 4. **Two-step model of Fermi arcs.** (a) Fermi surfaces of YBCO$_7$ for G-type AFM (red curves) and C-type AFM (blue curves) order. (b) Colored dots indicate relative intensity of spectral function due to coherence factor $w$, from Appendix III. (c) Fluctuating interlayer coupling smears the Fermi surfaces to fill in the space between the two magenta curves and between the two green curves [vertical shading].

## IV. TOPOLOGICAL NATURE OF DOMAIN WALLS

What is the physical meaning of the patch percolation transitions? In Ref. [2] it was shown that the various stripe energies could be understood as the energies of interacting topological defects – variously called antiphase boundaries (APBs), domain walls, or solitons[10]. That is, the percolation transition is a form of defect unbinding, related to the BKT transition[11] but involving extended defects rather than point vortices. In this case, the partition function can be approximately written as a product of a spin wave part and a topological defect (solitonic) term[10,11], $Z = Z_{sw}Z_{TD}$. Here we mainly discuss the latter contribution. Analytic results for solitons were derived in a weakly-coupled chain model, which corresponds to one-atom wide solitonic patches[10]. Our stripe patches differ in two ways. First, they are two Cu wide to account for AFM correlations. Second, since we are in the high soliton density limit, the different charge stripe periodicities account for fluctuating soliton interactions.

It was found[2] that the stripe patch energy per planar $CuO_2$ could be written as

$$E = \alpha/P_c + \beta/P_c^2 \quad (1)$$

with surface tension $\alpha/2a$, where $a$ is the lattice constant and APB interaction given by $\beta$. In $YBCO_6$ the APBs are only weakly charged, and $\alpha$ is approximately equal to the exchange constant J[2]]. Hence the Mott transition occurs on a J-scale, much lower than the Hubbard U scale. In contrast, in $YBCO_7$ the domain walls are charged, and screening by phonons leads to a much smaller value of $\alpha$, so unbinding occurs on the pseudogap temperature scale. We note that the attractive domain walls interaction ($\beta$ <0) can be interpreted as a screening effect, $\alpha \to \alpha/\varepsilon$, with dielectric constant $\varepsilon = 1 + P_0/P_c$, and

$$P_0 \approx -\beta/\alpha > 0.$$

While Eq. 1 is adequate for $YBCO_6$, there are small residual effects in $YBCO_7$ which lead to the Pc=4 stripe being the ground state[2].

The patch model is designed to provide a general picture of how excited states play a role at finite T in the physics of correlated electronic materials. The model is generic, designed to work with DFT results for any material, while improvements of the model are likely to be material-specific. Thus, in cuprates the stripe phases are related to topological defects – domain walls of AFMs – and improvements must take into account the topological nature of these walls. In contrast, in other materials (including LSCO) the phase competition may arise from octahedral tilts, with quite different defect physics.

By working with minimal fluctuations we expect accurate results at high T where fluctuations are first forming, while the proliferation of one type of fluctuation at low T gives an indication of grain growth. However, incorporating real grain growth into the model is difficult because one would have to incorporate the topological nature of domain walls – that they cannot begin or end without creating a new defect. Some consequences of grain growth are discussed in Appendix I.

Above the percolation crossover, the correlation length may get so small that no well-defined patches remain, but only a paramagnetic fluid of randomly oriented spins. This could, perhaps, describe the strange metal phase of cuprates. A possible model of such a phase is briefly discussed in Appendix IV.

## V. RELATION TO MANY-BODY PERTURBATION THEORY

A limitation of the patch model is that it is based on the number of low-energy phases found in a DFT calculation, which will only capture a small subset of possible phases. Hence, an important issue is, to what extent is this subset representative of the whole? Here we show that the present results are complementary to recent DFT + MBPT (many-body perturbation theory) calculations[23], which find a continuous distribution of competing phases. This complementarity lends further support to the results of both approaches.

Similar problems of many competing orders arise fairly often in condensed matter physics – including in CDWs[24], AFMs[25], FMs[26], and glasses[27] – and are often handled by mode-coupling theory. When too many modes try to soften at the same time, they frustrate any transition, so while one mode may win out at low-T, there is a broad range of T where long-range order is lost by the entropy gained by mixing many modes [as in Fig. 2(b) above]. A mode-coupling theory applied to cuprates[23,28] produces results broadly consistent with those of Figs. 2(a) and (b). This effect of mode-coupling theory is being incorporated in a variety of beyond-DFT codes[29] via a "Moriyaesque $\lambda$- factor". The multitude of phases found in YBCO by SCAN seem consistent with this mode-coupling interpretation.

We find that the DFT results and the mode coupling calculations mutually illuminate one another. Thus, the contrasting behavior between YBCO$_6$, Fig. 2(a), and YBCO$_7$, Fig. 2(b), is associated in mode coupling theory with a Mott-Slater transition. In the Mott phase the order is commensurate ($\pi,\pi$) Neel order independent of the Fermi surface shape and the correlation length grows rapidly (exponentially) at low T, whereas the Slater phase is characterized by many competing incommensurate orders dominated by Fermi surface nesting, leading to slow (power-law) growth of the correlation length.

Mode coupling leads to competition among a broad spread of states in momentum space, corresponding to a form of localization in real space, and the suggestion that this localization is due to local moment formation[28] is now confirmed by the DFT results[2], which find that all of the low energy magnetic and stripe phases are characterized by optimizing the value of the copper magnetic moment – i.e., by local moment formation. Moreover, this real-space localization is directly attributable to the appearance of solitons, forming a direct link between mode-coupling theory and various theories of topological defects.[9-11]

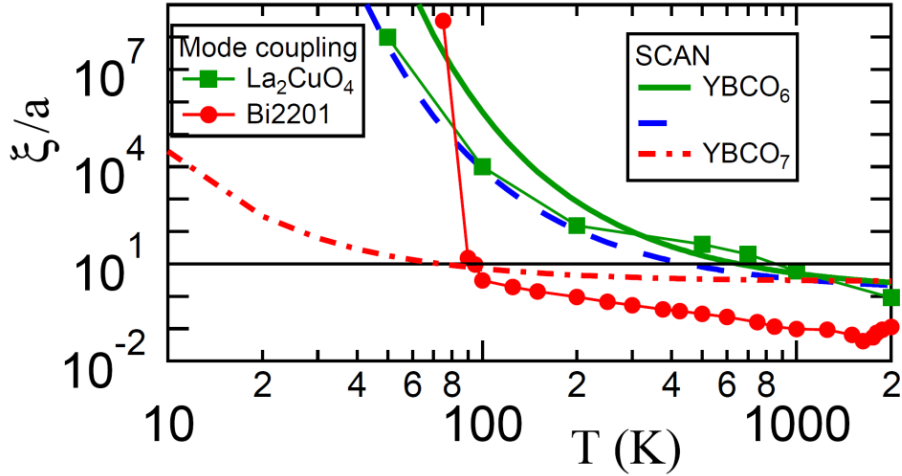

Figure 5. Correlation lengths in cuprates estimated by two techniques. Thin green curve with squares is the mode coupling result for undoped La$_2$CuO$_4$ in t-t'-t'' approximation, with t''=-t'/2 and t'/t = -0.13, taken from Fig.12(b) of Ref. [23], while the red curve with circles is the corresponding result for x=0.20 Bi2201 (Fig. 14). Thick solid lines represent the solitonic result Eq. 1 with Es = 110 (green solid line), 75 (blue dashed line), and 10 (red dot-dashed line) meV.

While mode coupling leads to the expectation that entropic effects will turn on near $T \sim J$ [28], the exact mechanism has been unclear. The present results suggest that this is due to solitonic deconfinement. For solitons there is a simple relationship between the AFM correlation length and the soliton energy Es:

$$\xi = W\exp\left[-\frac{\text{Es}}{\text{T}}\right]2^{1/2}, \tag{2}$$

where $W \approx a$ is the width of the domain wall[10]. Figure 5 compares the results of Eq. 2 (thick lines) with mode coupling results for undoped La$_2$CuO$_4$ (green squares) and x=0.20 doped Bi2201 (red circles). Since in Eq. 2 Es is a single soliton energy, we approximate it as $Es = \alpha$, from Eq. 1, where $\alpha$ = 110meV

(green solid line) for YBCO$_6$, and $\alpha$ = 10meV (red dot-dashed line) for the lowest energy stripes in YBCO$_7$.[2]

From Fig. 5 it is evident that Eq. (2) captures the Mott-Slater crossover: $\xi/a$~100 at 300K for YBCO$_6$, but only near 20K in YBCO$_7$. These results are in good agreement with the mode-coupling results, despite differences in detail. For instance, La$_2$CuO$_4$ has an inflection point near 300K where Fermi surface features become coherent, which is not captured by Eq. 2. While the results for undoped La$_2$CuO$_4$ are similar to those for YBCO$_6$, a better fit at low T is found with a smaller $E_s$=70meV, blue dashed curve. The Bi2201 results are similar to YBCO$_7$, (red dot-dashed curve). For the latter comparison we note two points: (1) the self-consistent mode coupling equation readily gives $\xi<a$, but it is not clear what this means experimentally, and we simply assume $\xi \leq a$ means uncorrelated. It is more significant that at T=100K $\xi/a$ <10 for both YBCO$_7$ and Bi2201, consistent with experiment.[30,31] (2) The sharp increase near 80K in Bi2201 is associated with the ground state becoming a single-$q$ CDW, and not the AFM. This is analogous to Eq. 1 describing only the G-AFM and not the $P_c$=4 bond-centered stripe ground state. The red dot-dashed line shows that the G-AFM order would finally develop a long-range correlation length, but only near 10K. The important role of the susceptibility $\chi$ in mode-coupling theory is analogous to Feynman's model of the roton minimum in helium, based on the dynamic structure factor $S(q,\omega) \sim Im(\chi)$.[9]

Since mode-coupling theory postulates a strong growth of entropy with T, it is interesting to see what the patch model predicts. Fig. 6(a) shows that the patch free energy is negative at high T, showing that the ground state is a spin liquid, as in BKT theory. At the same time, the Sommerfeld constant $\gamma$ = $C_V/T$, Fig. 6(b), shows a peak which sharpens and shifts toward T=0 with increased doping, suggestive of approach to a quantum critical point. This should be compared with Fig. 11 of Ref. [23]. These results are derived and further discussed in Appendix I.

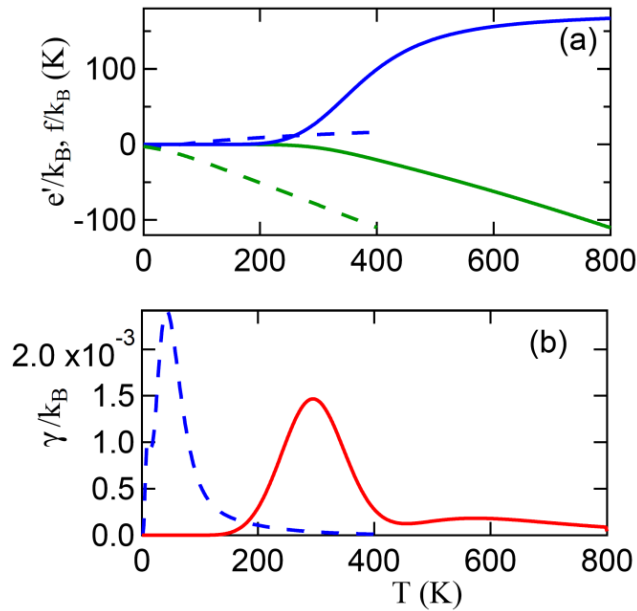

Figure 6. (a) Energy (blue lines) and free energy (green lines) per planar Cu for YBCO$_6$ (solid lines) and YBCO$_7$ (dashed lines). (b) Sommerfeld constant $\gamma$ for YBCO$_6$ (red solid line) and YBCO$_7$ (blue dashed line).

## VI. DISCUSSION

Intertwined order[1] is often taken to refer to an emergent supersymmetry, wherein a few orders (typically superconductivity, AFM, and/or CDW) are combined into a single multicomponent order parameter. In contrast, in YBCO our analysis indicates that the competing phases are related to topological defects, leading instead to a fluctuating stripe model.[32]

In addition to the topological defect partition function, there is also a spin contribution: as T is increased the spins will evolve from Ising to X-Y to Heisenberg, with increasing fluctuations. We believe this will not greatly affect the patch physics, as long as the magnetic correlation length is large. The spins will retain the underlying AFM correlations over the length scale of the patches, but with the spins aligned along tilted axes. This is analogous to the BKT problem, where the partition function separates into a magnetic part and a defect part.

While the patch model suggests a stripe ground state for $YBCO_7$, our model leaves out a number of effects which could cause some patches to persist at low temperatures, including disorder[33], possible nanoscale phase separation, and superconductivity. Also, it is not clear that the experiments are always in thermal equilibrium, as cooling too fast can lead to an excess of topological defects [Kibble-Zurek physics].

Could a patch model be relevant for the moment problem in other materials? It is notable that multiphase approaches can correctly model the Curie temperature of Fe and Ni as a percolation crossover,[34,35] and that mode coupling approaches have been applied to ferromagnets[25].

## VII. CONCLUSION

While DFT is a ground state theory, our modeling provides a scheme for obtaining a handle on the finite temperature properties of complex materials in terms of the statistical mechanics of the competing low-energy states. In particular, we show that the AFM order in $YBCO_6$ gives way to a spin-liquid phase at J-scale temperatures, which persists to U-scale temperatures, as expected for a Mott transition. Moreover, in $YBCO_7$ a fluctuating pseudogap phase persists to temperatures below the superconducting transition. Although we have not addressed superconducting phases, it is encouraging that the superconducting gap related patches observed in STM correspond to different local densities, as expected for stripe patches. We have further tested the present model by showing that it reasonably captures nematic effects in $YBCO_7$. The present approach should be applicable to a large variety of correlated materials where the competing phases are predominantly electronic. While domain walls in magnetic systems and dislocations in incommensurate density waves are obvious candidates, there are many systems composed of fairly rigid units – e.g., octahedra – which can easily tilt about various axes, and a patch model should be able to describe this competition.

## ACKNOWLEDGMENTS

The work at Tulane University was supported by the start-up funding from Tulane University, the Cypress Computational Cluster at Tulane, the DOE Energy Frontier Research Centers (development and applications of density functional theory): Center for Complex Materials from First Principles (CCM) under DE-SC0012575, and the National Energy Research Scientific Computing Center supercomputing center (DOE grant number DEAC02-05CH11231). The work at Northeastern University was supported by the U.S. DOE, Office of Science, Basic Energy Sciences grant number DE-FG02-07ER46352 (core research) and benefited from Northeastern University's Advanced Scientific Computation Center, the National Energy Research Scientific Computing Center supercomputing center (DOE grant number DEAC02-

05CH11231), and support (testing the efficacy of new functionals in diversely bonded materials) from the DOE Energy Frontier Research Centers: Center for Complex Materials from First Principles (CCM) under DE-SC0012575. We thank Hoang Tran, Carl Baribault, and Hideki Fujioka for their computational support at Tulane University.

**Appendix I: Entropy of Mixing**

While much of our thermodynamic extension of DFT is similar to earlier models[36], we focus on applying it to minimal fluctuations. Hence, some care must be taken in defining the entropy of the paramagnetic fluid (spin liquid). The average entropy should be the same on every site, $F=Nf$, where $f$ is the entropy per planar copper and $N$ the number of planar copper atoms. The free energy per site can be written $f = f_{av} - T s_m$, where $f_{av}$ is the average free energy of the individual phases, and $s_m$ is an additional entropy of mixing. Here we derive the free energy $f_{av}$, which is the probability that a planar copper atom is in phase $i$ times the free energy per planar copper atom, $f_{0i}$, summed over $i$, or $f_{av} = \sum_i a_i f_{0i}$, where $f_{0i} = F_i/n_i$, with $F_i = E_i - TS_i$, $E_i = n_i E_{0i}$, $S_i = k_B \ln(n_i)$. Then

$f_{av} = (Z/A)\underline{F}$,     (A1)

with $\underline{F} = \underline{E} - T\underline{S}$, and $\underline{X} = \sum_i z_i X_i$. The extra term $A/Z$, a weighted patch area, arises because the $Z_i$ describe the probability of exciting a finite-size fluctuation, whereas $f$ is a property of individual lattice sites. Thus, if we assume, following Eq. A1, that $f = (Z/A)F_p$, where $Z=\exp(-F_p/k_B T)$, we can find an explicit expression for the mixing entropy.

For patch maps, the Shannon entropy is

$$S_{\text{Sh}} = -\sum_i z_i \ln[z_i]. \qquad (A2)$$

Since $Z_i = \exp(-F_i/k_B T)$, Eq. A2 can be rewritten as

$F_p = \underline{F} - T S_m$,     (A3)

where $S_m = k_B S_{\text{Sh}}$ is the entropy of mixing. Then

$F_p = E_p - T S_p$,     (A4)

where $S_p = -\partial F_p/\partial T = \partial(k_B T \ln Z)/\partial T = \underline{S_p} + S_m$ and $E_p = F_p + T S_p = \underline{E}$. Finally, the heat capacity $C_{vp} = T \partial S_p/\partial T$.

Figure A1 shows our calculated values of $S_m$, $F_p$, and heat capacity for YBCO$_6$ (a,b) and YBCO$_7$ (c,d). In both cases, $E_p$ increases with T while $F_p$ decreases – as expected for an entropy-driven transition, predicted for cuprates[23] and similar to the BKT transition[11]. For this calculation we write $Z = Z_{sw} Z_{TD}$, and focus on the effect of $Z_{TD}$. Here we take $Z_{sw}$ as the G-AFM (at higher T it would include x-y and Heisenberg fluctuations). In YBCO$_6$ the results are consistent with the analysis of Fig. 2(a), with a peak in the heat capacity close to the Neel temperature. Note that the trends in heat capacity closely resemble those of BKT theory, including the fact that the peak in $C_{vp}$ lies above the transition temperature [see Fig. 13 in Ref. [11]]. Figure A2(a) shows the patch Sommerfeld constant $\gamma_p = C_{vp}/T$.

Figure A2(b) shows the average patch areas $A/Z$ for YBCO$_6$ (red solid line) and YBCO$_7$ (blue dashed line). The temperature dependence of $A/Z$ means that entropy s defined as $-\partial f/\partial T$ is not the same as $s' = S_p Z/A$.

Here we use the definition $x' = X_p Z/A$ for any thermodynamic quantity X, and the resulting thermodynamic quantities per planar Cu are plotted in Fig. A3, with the Sommerfeld constant $\gamma$ in Fig. 6(b). Note that when A/Z changes strongly with T it induces a dip in $\gamma$ compared to $\gamma_p$.

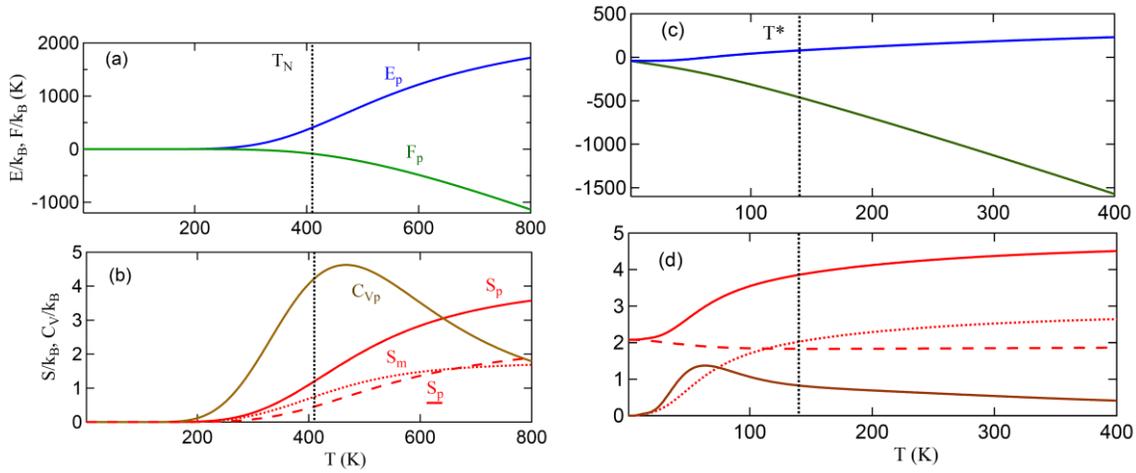

Figure A1. Patch thermodynamic functions for YBCO$_6$ (a,b) and YBCO$_7$ (c,d). (a,c) Patch energy (blue lines) and free energy (green lines). (b,d) Patch entropies S (red solid lines), $\underline{S_p}$ (red dashed lines), and $S_m$ (red dotted lines), and heat capacity (brown lines).

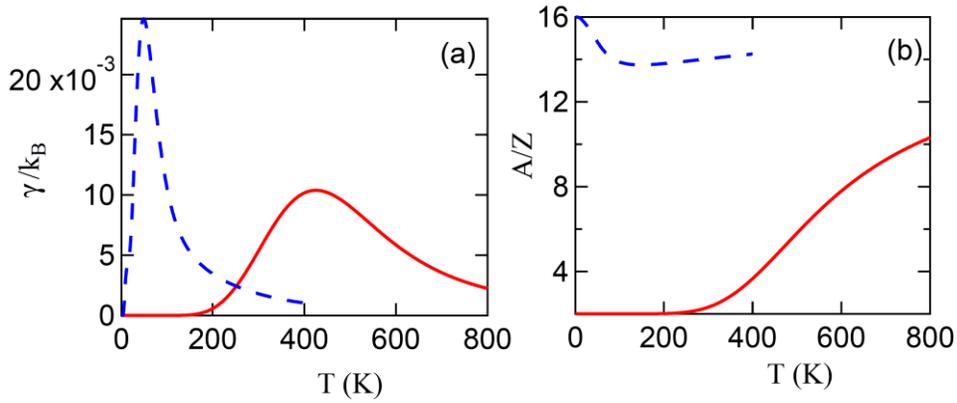

Figure A2. Patch Sommerfeld constant $\gamma_p$ (a) and average patch areas A/Z (b) for YBCO$_6$ (red solid line) and YBCO$_7$ (blue dashed line).

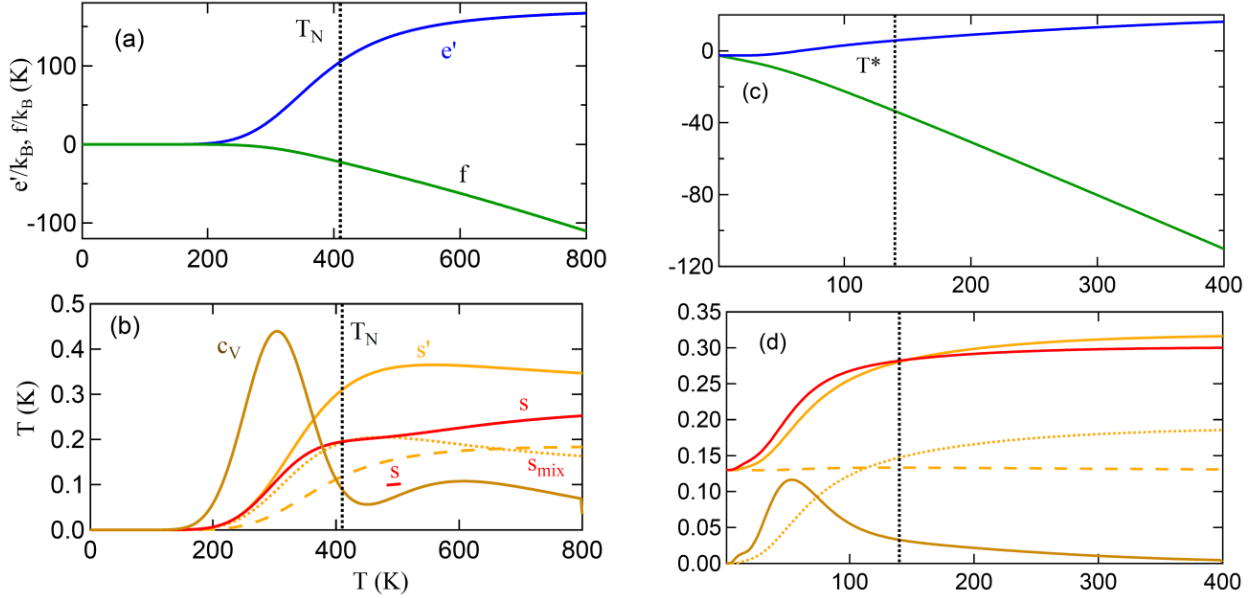

Figure A3. Thermodynamic functions per planar Cu for YBCO$_6$ (a,b) and YBCO$_7$ (c,d). (a,c) Energy e' (blue lines) and free energy f (green lines). (b,d) Entropies s (red solid lines), s' (orange solid lines), s'$_p$ (orange dashed lines), and s$_m$ (orange dotted lines), and heat capacity c$_v$ (brown lines).

We note that for YBCO$_7$ there is a residual entropy $S/k_B = 2.08 = \ln[8]$ of the P$_c$ = 4 stripe manifold at T=0, Fig. A1(d). While this is not true for the ground state, it is the correct result for the patch model: when a melt is cooled from high temperatures it becomes a polycrystalline solid unless special precautions are taken to form a single crystal. [Here 8=16/2: the unit cell contains 16 Cu atoms per CuO$_2$ layer, so the grains can nucleate on 16 inequivalent sites, while the AFM order has 2 inequivalent sublattices.] From Fig. A3(d), we see that this residual entropy is reduced by the number of particles per patch. However, at low temperatures the dominant patches do not simply proliferate, they also grow in size, further reducing the residual entropy, and this last effect is not accounted for in the patch model.

A second possible problem arises with the free energy *f* of YBCO$_6$. One would have expected that *f* first increases with *T*, then changes sign at a higher *T*, as in BKT theory. Again, there is too much entropy at low *T*. Allowing for growth of domains should also correct this feature.

**Appendix II: Fermi surface Nesting**

In section III.A we suggested that the multiplicity of low-lying patch periodicities is a consequence of building incommensurate CDWs from an array of commensurate stripes. Here we further explore the relation between the stripe phases seen at low doping and the CDW seen at higher doping, which is predicted in many models[37-40] to be associated with Fermi surface or hot-spot nesting. Indeed, since domain formation is usually entropically driven, we expect the energy per planar Cu to decrease monotonically with $P_c$. Thus, the appearance of a preferred $P_c$ is suggestive of new physics, which we propose is Fermi surface nesting – i.e., the CDWs are built from the various stripe patches, consistent with [38-40]. Since the effective exchange *J'* must decrease with doping, and long-range magnetic order will be lost for *T~J'*, we expect that, even at low doping the stripe magnetic order will onset at lower *T* than the charge order, while only short-range magnetic order may exist at higher doping, leaving a CDW behind. Both of these expectations are consistent with experiment. Also, the incommensurate CDWs found in STM experiments are modeled as commensurate patches separated by dislocations [13], with the incommensurate period arising as an average over many commensurate periods. This bears a striking resemblance to our patch maps in Fig. 1. Moreover, the importance of electron-phonon coupling[2] is consistent with CDW order in cuprates.[37,41] At the same time, Fermi surface nesting would seem like the best explanation for why we find a large variety of coexisting stripe periodicities, for the even-odd dichotomy, and for why a particular $P_c$=4 periodicity is singled out, as we here discuss.

For a stripe or CDW, the charge periodicity $P_c$ implies a set of preferred q-vectors $Q_{cm} \sim m/P_c$, where m is an integer with m<$P_c$, such that electronic bands with energy E(k) mix with those at E(k+$Q_{cm}$), with gaps opening where the bands cross. In general, these gaps do not fall at the Fermi level, and have little influence on the electronic energy. FS nesting arises when some $Q_{cm}$ approximately spans some part of the FS, so that the corresponding gap crosses the Fermi level, leading to a significant lowering of the electronic energy. By studying the theoretical dispersions, we find that the $P_c$=4 bond-centered stripe has a minigap that is nearly centered at $E_F$, for m=1, Fig. B1(a). Indeed, it is known that for many cuprates antinodal and hot-spot nesting are both optimal near q=0.25.

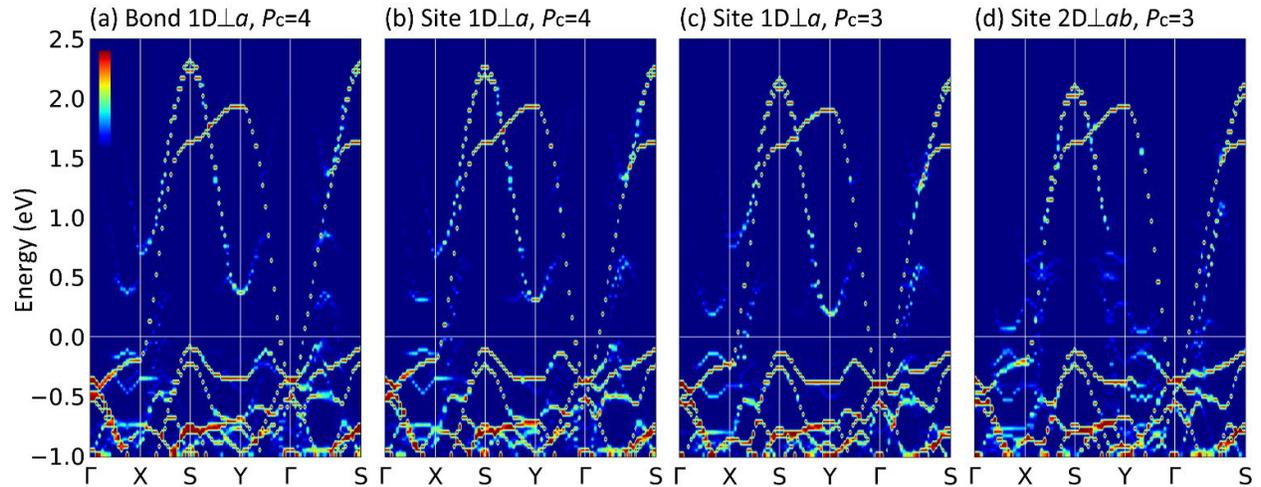

**Figure B1. Band structures of several YBCO$_7$ stripe phases.** (a) 1D bond-centered stripe phase with a charge stripe periodicity $P_C$ = 4. (b) 1D site-centered stripe with $P_C$ = 4. (c) 1D site-centered stripe with $P_C$ = 3. (d) 2D site-centered stripe with $P_C$ = 3 in both directions. The wavevector X/Y is at the first Brillouin zone boundary of lattice $a/b$. The Fermi levels are at zero.

Similar arguments explain the even-odd dichotomy. In addition to the FS gap for $P_c$=4, there is also a charge gap for $P_c$=8, m=2, so $Q_{c2} \sim 2/8$, and a magnetic gap for $P_c$=6 ($P_m$=12) at $Q_{m3}$=3/12. Indeed, one finds some gap at q=0.25 for every even $P_c$. In contrast, for $P_c$ odd, the gap is away from q=0.25: at 0.2 for $P_c$=5 and 0.28 for $P_c$=7, m=2. Finally, we note that superconductivity has been found to compete with both CDWs[23] and nematicity[32], further evidence for a close connection between the two phenomena.

Perhaps the strongest evidence that the stripes are CDW-driven is Fig. 6(b): the Sommerfeld constant $\gamma$ suggesting an approach to a quantum critical point. This same effect was found in Ref. [23], Fig. 11, but there it was clearly linked to the Van Hove singularity approaching the Fermi surface. This in turn is related to the CDW instability.

**Appendix III: Tight-binding model of YBCO**

In analyzing the c-axis dispersions of the various AFM phases, we find that the first-principles dispersions are consistent with the following tight-binding model. Including AFM order with gap $\Delta$ and interlayer coupling, the dispersion is

$E_\pm = E_{0p} \pm E_r$,

with

$E_r^2 = E_{0m}^2 + \Delta^2/4$,

$E_{0m} = -2t(c_x+c_y) - 2t'''(c_{2x}c_y + c_{2y}c_x) + E_{zG}$,

and

$E_{0p} = -4t'c_x c_y - 2t''(c_{2x}+c_{2y}) + E_{zC} - E_F$.

The interlayer couplings $E_{zi}$, $i = C, G$, are given by $E_{zi} = E_z \delta_{ij}$ for the j-AFM order, with

$E_z^2 = t_z^2 + t_z'^2 + 2t_z t_z' c_{2z} = t_{z1}^2 + t_{z2}^2 c_z^2$,

where $t_z$ refers to interlayer hopping across the Y-layer, and $t_z'$ across the CuO chains, and $t_{z1} = t_z - t_z'$, $t_{z2} = 2t_z t_z'$.

The band unfolding procedure introduces a structure factor weighting the intensities, consistent with the forms introduced for AFM and stripe orders in cuprates.[21] For the AFM orders, the resulting weights are

$w_\pm = (1 \pm E_{0m}/E_r)/2$.

Figure C1 shows first-principles SCAN results for the YBCO$_7$ Fermi surfaces, comparing nonmagnetic [NM] (a,b) with G-AFM (c,d) and C-AFM (e,f). The corresponding tight-binding model results are shown in Figure 4.

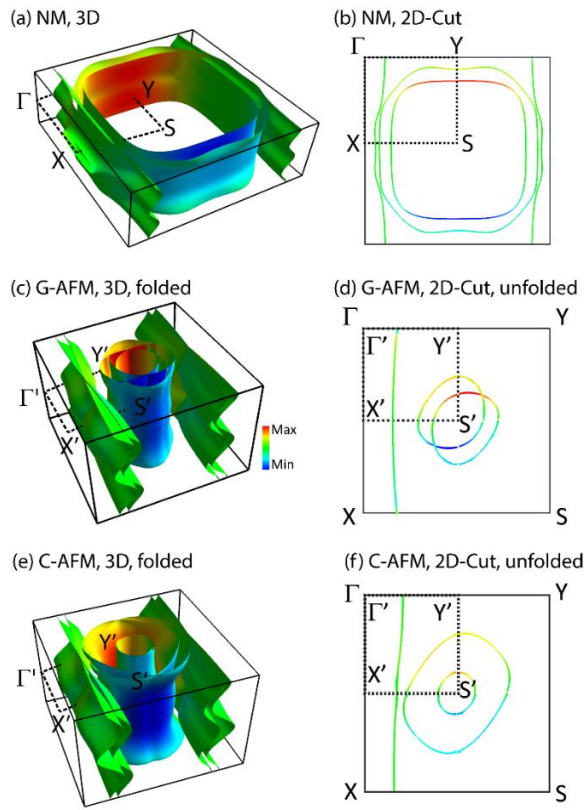

Figure C1. SCAN calculated Fermi surfaces for YBCO$_7$ showing full Fermi surfaces (a,c,e) and 2D cuts of the $\Gamma$-plane (???), for the nonmagnetic (a,b), G-AFM (c,d), and C-AFM (e,f) phases.

## Appendix IV: SQS and Paramagnetic Metals

**Table D1.** Relative ground state energies and local magnetic moments of YBCO$_7$ with paramagnetic (PM) ground state.

| Magnetic structure | Atom numbers and cell sizes | Local magnetic moment ($\mu_B$/Cu) | Energy (meV/CuO$_2$) |
|---|---|---|---|
| **YBCO$_7$** | | | |
| PM_52 | 52, irregular shape | 0.375~0.382, Avg=0.378 | 28.80 |
| PM_104 | 104, $2\sqrt{2} \times 2\sqrt{2} \times 1$ | 0.053~0.428, Avg=0.249 | 37.05 |
| PM_208 | 208, irregular shape | 0.231~0.434, Avg=0.348 | 27.43 |

We recall the special quasirandom structure (SQS)[42] which can be used to simulate the disorder caused by random spin directions, i.e., paramagnetism. Figure D1 shows the crystal structures of three SQS models, having supercells with 52, 104, and 208 atoms. Table D1 compares the total energies of these states to those of the nonmagnetic (NM) and G-AFM phases of YBCO7. While these paramagnetic phases have considerably lower energies than the NM phase, they are considerably higher in energy than the G-AFM phase, or many of the competing stripe phases[2]. Thus, while the SQS model shows that the NM phase lies too high in energy to play a role in the low-energy physics of cuprates, the assumption of completely random spin directions is too drastic when compared to the present model of short-range order. Figure D2 shows the evolution of the unfolded band structure from NM Fermi liquid to 52 atom paramagnet to G-AFM order. The larger structures have irregular shapes, and hence an unfolded band structure cannot be accurately determined.

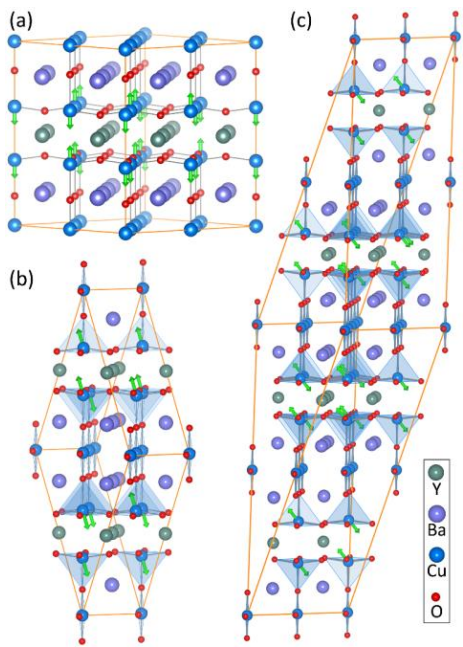

**Figure D1.** Special quasirandom structures used to simulate the paramagnetism induced by random spin orientations of the CuO$_2$ plane Cu ions: Crystal structures and magnetic configurations. **(a).** The SQS supercell is approximately $2\sqrt{2} \times 2\sqrt{2} \times 1$ of the unit cell, (b) The supercell with 52 atoms, and (c) the supercell with 208

atoms. In all frames the arrows denote the relative spin directions, since we did not determine the absolute spin directions by including the spin-orbit coupling.

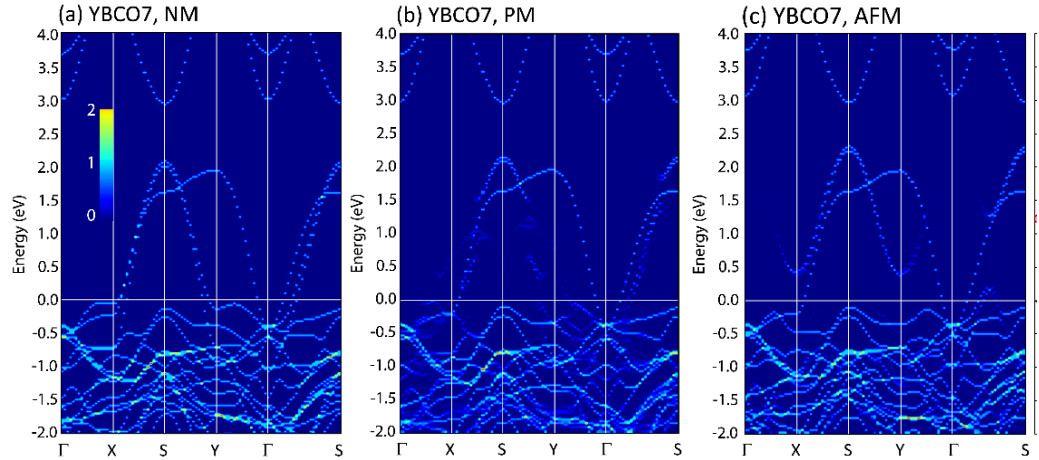

**Figure D2**. **Evolution of the effective band structures of YBCO$_7$ from the non-magnetic to the paramagnetic and to the antiferromagnetic spin structures.** (a) The non-spin-polarized results. (b) The paramagnetic results unfolded from the $2\sqrt{2} \times 2\sqrt{2} \times 1$ supercell. (c) The G-type antiferromagnetic results unfolded from the 2 × 2 × 1 supercell.